\documentclass[twocolumn,10pt]{revtex4}
\usepackage{graphicx}
\usepackage{amssymb,amsmath,amsfonts,amsthm}

\begin{document}

\title{Reheating temperature from the CMB}

\author{Jakub Mielczarek}
\email{jakub.mielczarek@uj.edu.pl}
\affiliation{Astronomical Observatory, Jagiellonian University, 30-244
Krak\'ow, Orla 171, Poland}

\begin{abstract}
In the recent paper by Mielczarek \emph{et al.} (JCAP {\bf 1007} (2010) 004)  an
idea of the method which can be used to put some constraint for the reheating phase was proposed. 
Another method of constraining the reheating temperature has been recently 
studied by Martin and Ringeval (Phys.\ Rev.\  D {\bf 82} (2010) 023511).
Both methods are based on observations of the cosmic microwave background (CMB) 
radiation. In this paper, we develop the idea introduced in this first article to put 
constraint on the reheating after the slow-roll inflation. We restrict our considerations 
to the case of a massive inflaton field. The method can be, however, easily extended 
to the different inflationary scenarios. As a main result, we derive an expression on the 
reheating temperature $T_{\text{RH}}$. Surprisingly, the obtained equation is independent 
on the unknown number of relativistic degrees of freedom $g_*$ produced during the reheating. 
Based on this equation and the WMAP 7 observations, we find $T_{\text{RH}}=3.5\cdot 10^6$ 
GeV, which is consistent with the current constraints. The relative uncertainty of the 
result is, however, very high and equal to $\sigma(T_{\text{RH}})/T_{\text{RH}} \approx 53$. 
As we show, this uncertainty will be significantly reduced with  future CMB experiments. 
\end{abstract}

\maketitle

\section{Introduction}
The reheating \cite{Kofman:1994rk, Kofman:1997yn} is a hypothetical process 
in which the inflaton field \cite{Guth:1980zm} is converted into the standard model 
particles. The mechanism of reheating is usually assumed to be a parametric production of  
particles \cite{Bassett:2005xm}. However, the considerations are purely speculative 
due to the lack of any possible empirical verification of the reheating phase.  Up 
to now, only some weak constrains on the basic parameters of reheating are available. 
In particular, the reheating temperature $T_{\text{RH}}$ can be constrained from the 
both sides. From bottom the constraint is given by the big bang nucleosynthesis (BBN), 
namely $T_{\text{RH}} \gtrsim 10$ MeV \cite{Hannestad:2004px, Kawasaki:2000en}. From 
the top, the constraint comes from the energy scale at the end of inflation  
$T_{\text{RH}} \lesssim 10^{16}$ GeV. 
Roughly 18 orders of magnitude remain to place the reheating temperature 
somewhere between. Worse, there is no observational window available at these 
energy scales. Such a window exists however at the energies of inflation. It is 
because the perturbations created during the inflation can be studied by its impact
on the cosmic microwave background (CMB) radiation and subsequently by the 
large scale structures (LSS). The method of constraining the reheating phase 
indirectly by the inflationary observational window was recently studied in Ref. 
\cite{Martin:2010kz}.  It was shown that it leads to the lower constraint on the 
reheating temperature $T_{\text{RH}} \gtrsim 6 $ TeV.
 
In this paper, we present an alternative method which can be used to fix at least 
some details of reheating based on observations of the cosmic microwave 
background (CMB) radiation. In comparison with the method presented in Ref. 
\cite{Martin:2010kz},  it will be possible not only to put a constraint on $T_{\text{RH}}$, 
but just fix its value.  The idea of the method was sketched in Ref. \cite{Mielczarek:2010ga}. 
It bases on the fact that the total increase of the scale factor from the observed part of 
inflation till now can be determined from the CMB. The number of $e$-foldings from 
the observed part of inflation till its end can be determined too. Based on this,  the 
$e$-folding number from the end of inflation till the recombination can be found. As 
we show, this can be used to determinate the reheating temperature. For simplicity, 
we assume the slow-roll inflation (described by a massive inflaton field), which is in good 
agreement with the CMB observations. After inflation, the inflaton field undergoes coherent 
oscillations at the bottom of a potential well.  The reheating takes place when the Hubble 
parameter $H$ falls to the value of the inflaton decay rate $\Gamma_{\phi}$. We assume 
that the reheating is instantaneous. After reheating, the standard radiation phase takes place. 
The evolution of radiation is assumed to be adiabatic. During the reheating, the effective 
number of relativistic species produced is given by $g_*$. The decay rate of the inflaton 
field can be related with the remaining two parameters $g_*$ and $T_{\text{RH}}$ by the Friedmann equation as follows 
\begin{equation}
\Gamma_{\phi}^2  \simeq \frac{8\pi}{3m^2_{\text{Pl}}} g_{*} \frac{\pi^2 T^4_{\text{RH}}}{30}, 
\label{Gamma}
\end{equation}
where $m_{\text{Pl}} = 1.22 \cdot 10^{19}$ GeV. Therefore, only two from the parameters 
of reheating $(\Gamma_{\phi}, T_{\text{RH}}, g_* )$ are independent.  In this paper, we show that the reheating 
temperature can be determined independently on the remaining two parameters. Up to now,
the constraints on $T_{\text{RH}}$ were dependent on the value of  $g_*$. However, in the equation
derived in this paper, the $g_*$ factors  surprisingly cancel out. Having $T_{\text{RH}}$, the decay
rate $\Gamma_{\phi}$ can be expressed in terms of $g_*$ only. 

The considerations presented in this paper are restricted to the simplest setup in order 
to capture the essence of the method. However, extension to the different 
inflationary scenarios and to the more detailed models of reheating can be done 
straightforwardly.   
  
\section{Method}

\begin{figure}[ht!]
\centering
\includegraphics[width=7cm,angle=0]{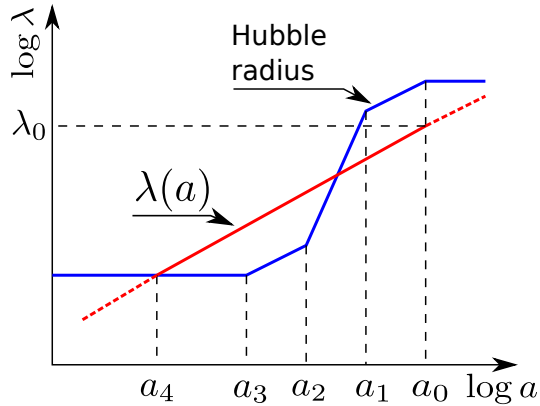}
\caption{Schematic evolution of the Hubble radius (blue line) for the 
standard cosmological scenario. The straight (red) line represents 
evolution of the physical length scale $\lambda(a)$, where 
$\lambda(a_0)=\lambda_0$.}
\label{Hubble}
\end{figure}

The main idea of the method can be understood by looking at Fig.  \ref{Hubble}.
In this figure we schematically present evolution of the Hubble radius $R_{\text{H}}:=1/H$, 
together with the evolution of an arbitrary physical length scale $\lambda$. The present 
value of this length scale is equal to $\lambda_0$, what we call the \emph{pivot scale}.  
The following values of the scale factor were distinguished: 
\\

$a_{0}$ -- the present value of the scale factor, we set $a_{0}=1$ for convenience. 

$a_{1}$ -- the scale factor at the end of the radiation era. Later, 
we set this to be a scale factor at the recombination which 
takes place soon after the end of the radiation era.
 
$a_{2}$ -- the scale factor at which the instantaneous reheating takes place 
(beginning of the radiation era).

$a_{3}$ -- the scale factor at the end of inflation. The inflation field starts to oscillate.

$a_{4}$ -- the scale factor at which the length scale of the present value 
$\lambda_0$ crossed the Hubble radius during inflation.  
\\

The total increase of the scale factor from $a_{4}$ to $a_{0}$ will be of particular 
importance. We call it $\Delta_{\text{tot}}$, which can be expressed as follows 
\begin{equation}
\Delta_{\text{tot}} = \prod_{i=0}^3 \Delta_i, \ \ \text{where} \ \  \Delta_i:=\frac{a_i}{a_{i+1}}.
\label{Dtoddef}
\end{equation}
So, if we know durations of the four stages between the $a_{4}$ and $a_{0}$, the 
$\Delta_{\text{tot}}$ can be determined. This is, however, practically impossible to 
obtain because we do not know details of the intermediate periods as $\Delta_2$ 
and $\Delta_1$.  Hopefully, there is an alternative method to determinate $\Delta_{\text{tot}}$,
which can be used to put constraints on $\Delta_2$ and $\Delta_1$. This method bases 
on the observation of the CMB radiation. In particular, on the measurements of the 
scalar power spectrum. The form of this spectrum is parameterized by  the function
\begin{equation}
\mathcal{P}_{\text{s}}(k) = A_{\text{s}} \left( \frac{k}{k_0}\right)^{n_{\text{s}}-1}. 
\label{powerspectrum}
\end{equation}
The $A_{\text{s}}$ is an amplitude and $n_{\text{s}}$ is a spectral index of 
the scalar perturbations. The $k_0$ is some arbitrary fixed scale called \emph{pivot number}.
We can relate it to the pivot scale $\lambda_0$ introduced earlier by  $\lambda_0=2\pi /k_0$.
In particular, the WMAP collaboration choice is  $k_0=0.002\ \text{Mpc}^{-1}$ (we also use this 
choice in this paper). For this value, the seven years of observations made by the 
WMAP satellite give the following values of the amplitude and spectral index of the scalar 
perturbations \cite{Komatsu:2010fb}
\begin{eqnarray}
A_{\text{s}} &=& 2.441^{+0.088}_{-0.092} \cdot 10^{-9},  \label{AsWMAP}\\ 
n_{\text{s}} &=& 0.963 \pm 0.012. \label{nsWMAP}
\end{eqnarray}
On the other hand, the well known prediction of the slow-roll inflation is
\begin{equation}
\mathcal{P}_{\text{s}}(k) = \underbrace{\frac{1}{\pi \epsilon} 
\left(\frac{H}{m_{\text{Pl}}} \right)^2}_{:=\mathcal{S}} \left( \frac{k}{aH}\right)^{n_{\text{s}}-1}, \label{spowersl}
\end{equation}
where $\epsilon$ is the so-called slow-roll parameter equal to
\begin{equation}
\epsilon = \frac{m^2_{\text{Pl}}}{4\pi} \frac{1}{\phi^2}  .
\end{equation}
For the considered massive slow-roll inflation $n_{\text{s}}=1-4\epsilon$. 
 
Let us now consider the power spectrum at the length scale $\lambda_0$ which 
corresponds to the pivot number $k_0$. From observation, an amplitude of the 
scalar perturbations at this scale is equal to $\mathcal{P}_{\text{s}}(k_0) = A_{\text{s}}$. 
On the other hand, this amplitude is formed when $k\simeq a H$. Therefore, for the mode 
$k_0$ we have $\mathcal{S}=A_{\text{s}}$. 

Cosmological evolution of the pivot scale $\lambda_0$ is given by 
\begin{equation}
\lambda(a)=\lambda_0 \frac{a}{a_0}.
\end{equation}
This relation is represented by the red line in Fig. \ref{Hubble}. The value of 
$\lambda$ was equal to the Hubble radius at $a_4$. Based on this, one can derive
\begin{eqnarray}
\Delta_{\text{tot}} = \frac{a_0}{a_4} = \frac{\lambda_0}{\lambda(a_4)} = \frac{H}{k_0},
\end{eqnarray}    
where $H$ is the value of the Hubble parameter when the $\lambda$ crossed 
the horizon during the inflation. In the second equality, we have used relation 
$\lambda(a) = \frac{2\pi}{k} \frac{a}{a_0}$, together with $k\simeq a H$ at the horizon 
crossing. Namely, $\lambda_0=\frac{2\pi}{k_0}$ and $\lambda(a_4) = \frac{2\pi}{H} \frac{1}{a_0}$,
where $a_0=1$. At the pivot scale, $\mathcal{S}=A_{\text{s}}$, so 
\begin{equation}
\frac{H}{m_{\text{Pl}}} = \sqrt{\pi \epsilon  A_{\text{s}}}.
\end{equation}
Expressing the $\epsilon$ from $n_{\text{s}}=1-4\epsilon$, we find
\begin{equation}
\Delta_{\text{tot}} = \frac{1}{2} \frac{m_{\text{Pl}}}{k_0} \sqrt{\pi (1-n_{\text{s}})A_{\text{s}}}. 
\label{Dtotexpr}
\end{equation} 
The essential conclusion derived from this equation is that:
\emph{Based on the CMB observations, one can determinate the total increase of 
the scale factor from the observed moment of inflation till now.}
In principle, from the WMAP 7 observations we determinate 
\begin{equation}
\Delta_{\text{tot}} =  (8.0 \pm 1.5) \cdot 10^{51}. 
\end{equation} 

\section{Inflation and reheating}

The increase of a scale factor during the part of inflation  from $a_4$ to $a_3$ 
is given by 
\begin{equation}
\Delta_3= e^{N_{\text{obs}}}, \label{D3}
\end{equation}
where $N_{\text{obs}}$ is the $e$-folding number, which can be expressed as follows 
\begin{eqnarray}
N_{\text{obs}} &\simeq&-\frac{8\pi}{m^2_{\text{Pl}}}\int_{\phi_{\text{obs}}}^0 \frac{V(\phi)}{V'(\phi)} d\phi   
\nonumber \\
&=& 2\pi \frac{\phi^2_{\text{obs}}}{m^2_{\text{Pl}}} =\frac{2}{1-n_{\text{s}}}. 
\label{expNobs}
\end{eqnarray}
We have used here $V(\phi)=\frac{m^2}{2}\phi^2$ and defined $\phi_{\text{obs}}=\phi(a_4)$.  
In particular, based on the WMAP 7 data one can find $N_{\text{obs}}=54 \pm 18$. The 
uncertainty is high because of the strong sensitivity on the uncertainty of the spectral 
index $n_{\text{s}}$. This will later propagate to the uncertainty of the reheating temperature.  
As we show in Sec. \ref{Forecasting},  the uncertainty  of $N_{\text{obs}}$ can be significantly reduced with the future CMB experiments.

During the slow-roll inflation, evolution of the inflaton field
$\phi$ is well approximated by 
\begin{equation}
\phi(t)=\phi_{\text{max}} - \frac{m\ m_{\text{Pl}}}{\sqrt{12\pi}} t. 
\end{equation}       
From comparison with the numerical results, it can be seen 
that this approximation holds till the end of inflation, when $\phi \approx 0$. 
Therefore, the kinetic term 
\begin{equation}
\frac{\dot{\phi}^2}{2} \simeq \frac{m^2m^2_{\text{Pl}}}{24 \pi},
\end{equation}
is approximately constant during the inflation. This contribution to the total energy density 
is, however, dominated by the potential part during the slow-roll inflation. At the 
end of inflation, the contribution from the potential part falls to zero ($V(\phi=0)=0$), and 
the kinetic term dominates. One can therefore estimate that, at the end of inflation, the 
energy density is given by 
\begin{equation}
\rho(a_3) \simeq \frac{m^2m^2_{\text{Pl}}}{24 \pi}. \label{rhoa3}
\end{equation} 
A validity of this approximation was confirmed by the 
numerical computations. 

After inflation, the field starts to oscillate at the bottom of 
the potential well. During this evolution, the energy density 
drops as in the matter dominated universe \cite{Starobinsky:1980te}
\begin{equation}
\rho(a)  \simeq \rho(a_3)\left(\frac{a_3}{a} \right)^3.  \label{rhoaco}
\end{equation}
This evolution holds till $a_2$, when $H \approx \Gamma_{\phi}$ and the 
reheating takes place. Then, the energy density
\begin{equation}
\rho(a_2) = g_{*} \frac{\pi^2 T^4_{\text{RH}}}{30}, \label{rhoa2}
\end{equation} 
here $g_{*}=g( T_{\text{RH}})$ is the number of ultrarelativistic degrees of freedom 
generated during the reheating, where $g(T)$ is defined as follows
\begin{equation}
g=\sum_{\text{boson}} g_{\text{B}}+\frac{7}{8}\sum_{\text{fermion}} g_{\text{F}}. 
\end{equation} 
In particular, for Glashow-Weinberg-Salam (GWS) model 
$SU(2)_L \otimes U_Y(1) \otimes SU_{c}(3)$, we have $g=106.75$. 
Therefore one may expect that $g_* \geq 106.75$ if the temperature of 
reheating is greater than the electroweak energy scale, $T_{\text{RH}} \gtrsim 300$ GeV.

\begin{figure}[ht!]
\centering
\includegraphics[width=8cm,angle=0]{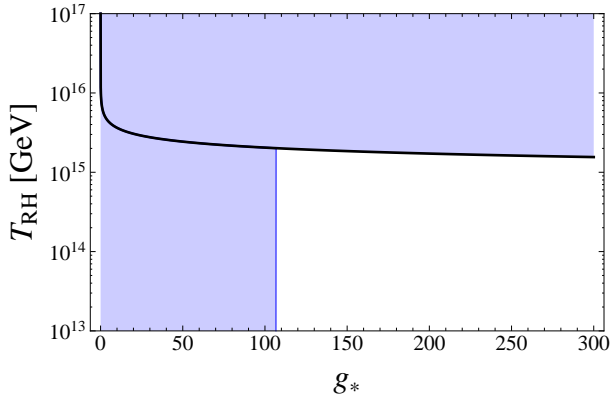}
\caption{Section of the reheating parameters space $(T_{\text{RH}},g_{*})$. 
The shadowed region is excluded by Eq.  \ref{condition1} and $g_* > 106.75$ 
valid for $T_{\text{RH}} \gtrsim 300$ GeV. The thick line represents  
$T_{\text{RH}} = g_*^{-1/4} \cdot  6.5 \cdot 10^{15} \ \text{GeV}$. }
\label{TRH1}
\end{figure}

Based on (\ref{rhoaco}) we have
\begin{equation}
\Delta_2= \frac{a_2}{a_3} = \sqrt[3]{\frac{\rho(a_3) }{\rho(a_2) }},
\end{equation}
and applying (\ref{rhoa3}) and (\ref{rhoa2}) we derive 
\begin{equation}
\Delta_2=\frac{1}{\pi} \left(\frac{5}{4}\cdot \frac{m^2m^2_{\text{Pl}}}{g_{*} T^4_{\text{RH}}} \right)^{1/3}. \label{D2}
\end{equation}
This result will be useful in the subsequent section when deriving the expression 
on $T_{\text{RH}}$. However, before we proceed to this issue we can see what we  
already can say about the reheating temperature. Let us notice that  the following 
condition $\rho(a_2)  \leq \rho(a_3)$ must be  fulfilled. Energy scale of reheating cannot
be higher than energy at the end of inflation.  In order to use this constraint one has 
to firstly determinate inflaton mass in Eq. (\ref{rhoa3}). It can be done by noticing 
that the Friedmann equation reduces to 
\begin{equation}
H^2 \simeq \frac{8\pi}{3m_{\text{Pl}}^2} \frac{1}{2} m^2 \phi^2
\end{equation}
in the slow-roll regime ($\epsilon \ll 1$). Based on this and condition $\mathcal{S}=A_{\text{s}}$,
together with $n_{\text{s}}=1-4\epsilon$, one can derive  \cite{Mielczarek:2010ga}
 \begin{equation}
m = m_{\text{Pl}} \frac{1}{4} \sqrt{3\pi A_{\text{s}}} (1-n_{\text{s}}). \label{mexpr}
\end{equation}
Applying this expression to the WMAP 7 results we obtain
\begin{eqnarray}
m &=& (1.4\pm0.5)\cdot 10^{-6}m_{\text{Pl}} \nonumber  \\
   &=&  (1.7\pm0.6) \cdot 10^{13} \text{GeV}.
\end{eqnarray}
With use of this value, the condition $\rho(a_2)  \leq \rho(a_3)$ reduces to 
\begin{equation}
g_*^{1/4} T_{\text{RH}} \leq 6.5 \cdot 10^{15} \ \text{GeV}. \label{condition1}
\end{equation} 
Based on this, part of the parameter space $(g_*, T_{\text{RH}} )$ can be excluded.
It was shown as the shadowed region above the thick line in Fig. \ref{TRH1}.
As we mentioned earlier, if  $T_{\text{RH}} \gtrsim 300$ GeV then  $g_* \geq 106.75$.
This constraint excludes another part of the parameter space. This was represented 
in Fig. \ref{TRH1} as the shadowed region constrained by the vertical line. Based 
on the above constraints, one can conclude that
\begin{equation}
T_{\text{RH}} \leq 2.0 \cdot 10^{15} \ \text{GeV}.
\end{equation} 
This corresponds to the inflationary bound on the reheating temperature. 

\section{Reheating temperature}

After reheating, the Universe is filled by the relativistic plasma.  
Expansion of this relativistic gas is assumed to be adiabatic 
and the masses of particles are neglected. The adiabatic 
approximation is valid until the entropy transfer between the 
radiation and other components can be neglected. 
In turn, this second approximation is valid if the 
temperature is much higher than the masses of the 
particles. Then, $dS=0$, which implies $s a^3=$ const, where the 
entropy density $s$ of radiation is given by 
\begin{equation}
s = \frac{2\pi^2}{45} g T^3.
\end{equation}
Based on this one can derive expression on the increase of the 
scale factor from reheating till the recombination
\begin{equation}
\frac{a_1}{a_2}=\frac{T_2}{T_1} \cdot \left(\frac{g_2}{g_1} \right)^{1/3}.
\end{equation} 
We have $T_2=T_{\text{RH}}$ and  $T_1$ is equal to the recombination 
temperature $T_{\text{rec}}$. During recombination $g_{1}=g_{\gamma}=2$ 
and during reheating $g_{2}=g_*$,  therefore 
\begin{equation}
\Delta_1= \frac{T_{\text{RH}}}{T_{\text{rec}}} \cdot \left(\frac{g_{*}}{2} \right)^{1/3}.  \label{D1}
\end{equation}
Finally, increase of the scale factor from recombination till now is given by    
\begin{equation}
\Delta_0 =1+z_{\text{rec}}, \label{D0}
\end{equation}
where $z_{\text{rec}}$ is the recombination redshift which can be determined from
the CMB observations. It is worth mentioning that an intermediate stage other than 
recombination can be used here. In particular, the equilibrium point (end of the radiation 
epoch, where $\rho_{\text{rad}}=\rho_{\text{mat}}$) can be chosen. The corresponding 
value of redshift can be also determined from the CMB observations.  

\begin{figure}[ht!]
\centering
\includegraphics[width=8cm,angle=0]{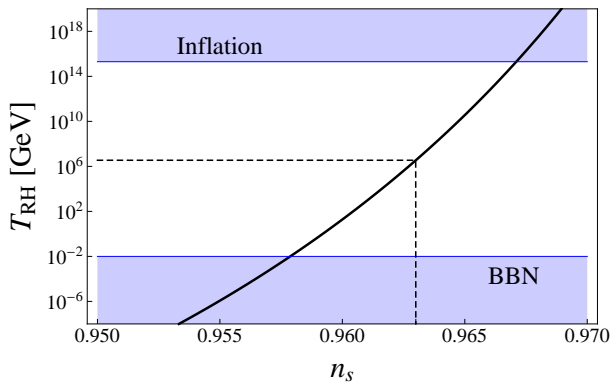}
\caption{The thick line represents Eq. \ref{TRHfinal} with $A_{\text{s}}= 2.441 \cdot 10^{-9}$. 
Dashed lines corresponds to $n_{\text{s}}=0.963$, taken from the WMAP 7 observations, 
what leads to $T_{\text{RH}}= 3.5\cdot 10^6\ \text{GeV}$. The shadowed regions are excluded 
by the inflationary and BBN constraints.}
\label{TRH2}
\end{figure}

At this point, we have all required to find the expression on $T_{\text{RH}}$.
We have found all $\Delta_i$ and $\Delta_{\text{tot}}$.   Based on (\ref{Dtoddef}),
the following relation is fulfilled
\begin{equation}
\Delta_{\text{tot}} =\Delta_3\Delta_2\Delta_1\Delta_0.
\end{equation}
Inserting  (\ref{D3}), (\ref{D2}), (\ref{D1}) and  (\ref{D0}) we obtain
\begin{equation}
\Delta_{\text{tot}} = e^{N_{\text{obs}}} \frac{1}{\pi} \left(\frac{5}{4}\cdot 
\frac{m^2m^2_{\text{Pl}}}{g_{*} T^4_{\text{RH}}} \right)^{1/3} 
\frac{T_{\text{RH}}}{T_{\text{CMB}}}  \left(\frac{g_{*}}{2} \right)^{1/3},
\end{equation}
where we have used $T_{\text{rec}}=T_{\text{CMB}}(1+z_{\text{rec}})$. 
The important observation is that $g_*$ factors cancel out. This is 
crucial, because the expression on $T_{\text{RH}}$ will be free from the 
dependence on the unknown $g_*$ parameter. With use of  (\ref{Dtotexpr}), (\ref{expNobs}) 
and (\ref{mexpr}), the above  equation can be rewritten into the following form  
\begin{equation}
T_{\text{RH}}=\frac{15\ m_{\text{Pl}}}{16 \cdot \pi^{7/2}}
\sqrt{\frac{1-n_{\text{s}}}{A_{\text{s}}}}
\left(\frac{k_0}{T_{\text{CMB}}}\right)^3  \exp \left\{ \frac{6}{1-n_{\text{s}}} \right\}. 
\label{TRHeq}
\end{equation}
This equation is a main result of this paper. Taking the constant parameters  
$T_{\text{CMB}}=2.725\ \text{K}=2.348 \cdot 10^{-4}$ eV and $k_0=0.002\ \text{Mpc}^{-1}$  
(and reexpressing units: Mpc${}^{-1}=6.39\cdot 10^{-30}$ eV) one can rederive 
Eq. (\ref{TRHeq}) to the practical form
\begin{equation}
T_{\text{RH}}= 3.36 \cdot 10^{-68}  \sqrt{\frac{1-n_{\text{s}}}{A_{\text{s}}}}  
\exp \left\{ \frac{6}{1-n_{\text{s}}} \right\}\  \text{GeV}. \label{TRHfinal}
\end{equation}
In Fig.  \ref{TRH2} we show relation (\ref{TRHfinal}) as a function 
of the spectral index $n_{\text{s}}$.  We also mark the regions excluded
from the inflationary constraint and the BBN constraint.
For the data from the WMAP 7 observations, Eq. (\ref{TRHfinal}) leads to
\begin{equation}
T_{\text{RH}}= 3.5\cdot 10^6\ \text{GeV}.
\end{equation}
The relative uncertainty of this result is 
\begin{equation}
\frac{\sigma(T_{\text{RH}})}{T_{\text{RH}}} \approx  53.
\label{uncTRH}
\end{equation}
Here  a first order Taylor expansion was applied when calculating propagation of uncertainties:
\begin{equation}
\sigma(T_{\text{RH}}) \approx \sqrt{ \left(\frac{\partial T_{\text{RH}}}{\partial n_{\text{s}}} \right)^2\sigma^2(n_{\text{s}})
+\left(\frac{\partial T_{\text{RH}}}{\partial A_{\text{s}}} \right)^2\sigma^2(A_{\text{s}})}.
\end{equation}
However, due to the strong (exponential) dependence of $T_{\text{RH}}$ on $n_{\text{s}}$, 
the applied linear approximation may turn out to be insufficient.  Therefore, one can expect 
greater uncertainty of $T_{\text{RH}}$ than obtained here. Future studies need to address 
this issue. The high relative uncertainty (\ref{uncTRH}) is mainly a result of the weakly determined 
value of $N_{\text{obs}}$, which is a function of $n_{\text{s}}$. In the next section, we will 
examine how this uncertainty can be reduced with the future CMB experiments. 

As it was discussed in Refs. \cite{Nakayama:2008wy, Kuroyanagi:2009br}, if 
$T_{\text{RH}} \sim 10^{6-9}\ \text{GeV}$, then it may be possible to measure 
$T_{\text{RH}}$ with the planned space-based laser interferometer experiments 
such as the Big Bang Observer. The value $T_{\text{RH}}= 3.5\cdot 10^6\ \text{GeV}$ 
obtained here fulfills this condition. Our prediction has therefore chance to be 
verified in future. 

Furthermore, based on Eq. (\ref{Gamma}), one can express the inflaton decay rate 
$\Gamma_{\phi}$ in terms of $g_*$:
\begin{equation}
\Gamma_{\phi}\simeq  1.7 \cdot  10^{-6}\ \sqrt{g_*}\ \text{GeV},
\end{equation}
where the previously derived value of $T_{\text{RH}}$ was used. 
It is reasonable to expect that $\Gamma_{\phi} = \alpha \cdot m$,
which comes from the Heisenberg uncertainty relation. The $\alpha$ 
is a dimensionless parameter.   With use of the inflaton mass found 
earlier we find
\begin{equation}
\alpha \simeq  10^{-19}\ \sqrt{g_*} .
\end{equation}
Based on this result, one can deduce that the inflaton decays into the very light
particles comparing with its mass. This is also the reason why the reheating 
takes place at the relatively low energies. It becomes unstable only when 
the sufficiently low energies are reached. However, the origin of this low vale 
of decay rate $\Gamma_{\phi}$ cannot be understood without a deeper understanding
of the inflationary cosmology. 

\section{Forecasting} \label{Forecasting}

As we have shown in the previous section, the value of $T_{\text{RH}}$ is 
strongly dependent on $n_{\text{s}}$. Therefore, the method presented can 
be used effectively only if the value of $n_{\text{s}}$ is determined with high
precision. The value of $n_{\text{s}}$ from the WMAP 7 observations 
is not determined sufficiently precise to obtain a strong prediction concerning 
the reheating temperature.  However, it may change if the new observational 
data will be available. In this section, we predict how the uncertainty on 
$T_{\text{RH}}$ will be reduced with the future CMB experiments. In particular, 
we consider the Planck satellite \cite{:2006uk} experiment which is currently on 
the stage of collecting data. We consider the ACTPol \cite{Niemack:2010wz} 
ground-based experiment which is under construction at present.  
We also consider the planned CMBPol \cite{Dunkley:2008am} satellite experiment.

The uncertainty of $T_{\text{RH}}$ comes mainly from $n_{\text{s}}$, therefore, 
in the considerations we fix the value of $A_{\text{s}}$. Following
Ref. \cite{Galli:2010it}, the expected uncertainties of $n_{\text{s}}$ from the 
mentioned CMB experiments are the following 
\begin{equation}
\sigma(n_{\text{s}}) = \left\{ 
\begin{array}{cc}   
0.0031 & \text{Planck}\\
0.0021 & \text{Planck+ACTPol}.\\
0.0014 & \text{CMBPol}
\end{array}   \right.
\label{errors}
\end{equation}
Based  on this, let us first see the resulting uncertainties of the
$e$-folding number $N_{\text{obs}}$. We find
\begin{equation}
\sigma(N_{\text{obs}}) = \left\{ 
\begin{array}{cc}   
4.5 & \text{Planck}\\
3.1 & \text{Planck+ACTPol}.\\
2.0 & \text{CMBPol}
\end{array}   \right.
\end{equation}
This significant reduction of the uncertainty of $N_{\text{obs}}$ (with respect to the
WMAP 7 results) will be crucial for determining the reheating temperature.  Based on (\ref{TRHfinal})  
with  (\ref{errors}), we forecast
\begin{equation}
\frac{\sigma(T_{\text{RH}})}{T_{\text{RH}}} = \left\{ 
\begin{array}{cc}   
13.5 & \text{Planck}\\
9.2 & \text{Planck+ACTPol}.\\
6.1 & \text{CMBPol}
\end{array}   \right. 
\end{equation}
Here, the values of the parameters $n_{\text{s}}$ and $A_{\text{s}}$ were set to be those 
obtained from the WMAP 7 observations. 

In order to have $\sigma(T_{\text{RH}})/T_{\text{RH}}$ smaller than unity,
the uncertainty of $n_{\text{s}}$ should be reduced by 2 orders of magnitude 
with respect to the WMAP 7 results. At present, there is however no experiment 
planned to reach such sensitivity. The uncertainty may be nevertheless additionally
reduced by combining data from the different available experiments.  This is 
possible because of the angular scale dependent sensitivity of the CMB
experiments. In particular, the ground-based experiments can provide much 
better data of the CMB polarization at the small angular scales (high multipoles) 
than the space-based experiments can. 

\section{Summary}

In this paper, we have developed a new method of constraining the reheating
phase after inflation. The method bases on the observations of the cosmic 
microwave background radiation.  In particular, the fact that the total increase 
of the scale factor from the observed part of inflation till now can be determined 
is used.  Based on this, we have found the expression on the reheating temperature.
The expression is free from the dependence on the unknown $g_*$ parameter.  
With use of the WMAP 7 results, we have determined  $T_{\text{RH}}=3.5\cdot 10^3$ TeV.
The relative uncertainty of this result is  equal to $\sigma(T_{\text{RH}})/T_{\text{RH}} 
\approx 53$. This high uncertainty can be, however, significantly reduced with the 
future CMB data. One can expect the reheating temperature to be quite precisely 
determined (reaching  $\sigma(T_{\text{RH}})/T_{\text{RH}} =\mathcal{O}(1)$) 
within the present decade. 

The value of reheating temperature determined in this paper is consistent with 
the known bounds,  in particular, with the lower bound $T_{\text{RH}} \gtrsim 6 $ TeV  
recently found in Ref. \cite{Martin:2010kz}. It is also interesting to note that within 
the supersymmetric extension of the standard model, the upper bound on the 
reheating temperature exists $T_{\text{RH}} \lesssim 10^4$ TeV (see Refs. 
\cite{Khlopov1984, Khlopovbook, Kallosh:1999jj}).  Our result is also in agreement 
within this condition. Finally, it is worth mentioning that the low value of the reheating 
temperature, as determined here, can have interesting implications on the phenomenology
of primordial black holes \cite{Khlopov:2004tn}.    

\section*{Acknowledgements}

Author would like to thank to Micha{\l} Ostrowski for helpful comments and suggestions.
JM has been supported  by Polish Ministry of Science and Higher Education 
Grant N N203 386437 and by the Foundation of Polish Science.

\end{document}